\documentclass[aps,twocolumn,prl,floatfix,lengthcheck,citeautoscript,superscriptaddress,longbibliography]{revtex4-2}

\usepackage{amssymb,amsfonts,amsmath,amsthm}
\usepackage{graphicx}%esto va sin el DVIPS!
\usepackage[english]{babel}
\usepackage[latin1]{inputenc}
\usepackage[pdftex,colorlinks=true,pdfstartview=FitH,linkcolor=blue,citecolor=blue,urlcolor=blue]{hyperref}

\usepackage{bm}
\usepackage{float}
\usepackage{mathtools}
\usepackage{color,soul}
\usepackage{times}
\usepackage{upgreek}
\usepackage{MnSymbol}
\usepackage{verbatim}
\usepackage[bottom]{footmisc}
\usepackage{bbold}

\usepackage[dvipsnames,svgnames,table]{xcolor}
\usepackage{hyperref}
\usepackage[english]{babel}
\usepackage[caption=false]{subfig}

\hypersetup{
	pdfstartview={FitH},
	colorlinks=true,
	linkcolor=NavyBlue,
	citecolor=Blue,
	filecolor=NavyBlue,
	urlcolor=NavyBlue
}

\newcommand{\bea}{\begin{eqnarray}}
	\newcommand{\eea}{\end{eqnarray}}
\newcommand{\be}{\begin{equation}}
	\newcommand{\ee}{\end{equation}}

\newcommand{\gone}{g^{(1)}}
\newcommand{\Omctc}{\nu_\mathrm{CTC}}
\newcommand{\Omph}{\nu_\mathrm{M}}
\newcommand{\psip}{\tilde{\psi}_+}
\newcommand{\psim}{\tilde{\psi}_-}
\newcommand{\DR}{\Delta\nu_\mathrm{R}}
\newcommand{\DvR}{\Delta\nu_\mathrm{R}}
\begin{document}
	
	\title{Coherent Control of a Polariton Continuous Time Crystal}
	
	\author{I. Carraro-Haddad}
	\email[Corresponding author: ]{ignacio.carraro@ib.edu.ar}
	\affiliation{Centro At{\'{o}}mico Bariloche and Instituto Balseiro,
		Comisi\'on Nacional de Energ\'{\i}a At\'omica (CNEA)--Universidad Nacional de Cuyo (UNCUYO), 8400 Bariloche, Argentina.}
	\affiliation{Instituto de Nanociencia y Nanotecnolog\'{i}a (INN-Bariloche), Consejo Nacional de Investigaciones Cient\'{\i}ficas y T\'ecnicas (CONICET), Argentina.}
	
	\author{G.~Usaj}
	\affiliation{Centro At{\'{o}}mico Bariloche and Instituto Balseiro,
		Comisi\'on Nacional de Energ\'{\i}a At\'omica (CNEA)--Universidad Nacional de Cuyo (UNCUYO), 8400 Bariloche, Argentina.}
	\affiliation{Instituto de Nanociencia y Nanotecnolog\'{i}a (INN-Bariloche), Consejo Nacional de Investigaciones Cient\'{\i}ficas y T\'ecnicas (CONICET), Argentina.}
	
	\author{A.~Fainstein}
	\email[Corresponding author: ]{alejandro.fainstein@ib.edu.ar}
	\affiliation{Centro At{\'{o}}mico Bariloche and Instituto Balseiro,
		Comisi\'on Nacional de Energ\'{\i}a At\'omica (CNEA)--Universidad Nacional de Cuyo (UNCUYO), 8400 Bariloche, Argentina.}
	\affiliation{Instituto de Nanociencia y Nanotecnolog\'{i}a (INN-Bariloche), Consejo Nacional de Investigaciones Cient\'{\i}ficas y T\'ecnicas (CONICET), Argentina.}
	
	\author{K. Biermann}
	\affiliation{Paul-Drude-Institut f\"{u}r Festk\"{o}rperelektronik, Leibniz-Institut im Forschungsverbund Berlin e.V., Hausvogteiplatz 5-7,\\ 10117 Berlin, Germany.}
	
	\author{P.~V. Santos}
	\affiliation{Paul-Drude-Institut f\"{u}r Festk\"{o}rperelektronik, Leibniz-Institut im Forschungsverbund Berlin e.V., Hausvogteiplatz 5-7,\\ 10117 Berlin, Germany.}
	
	\author{A.~S. Kuznetsov}
	\email[Corresponding author: ]
	{kuznetsov@pdi-berlin.de}
	\affiliation{Paul-Drude-Institut f\"{u}r Festk\"{o}rperelektronik, Leibniz-Institut im Forschungsverbund Berlin e.V., Hausvogteiplatz 5-7,\\ 10117 Berlin, Germany.}
	
	\date{\today}
	
	\begin{abstract}
		{
			Spontaneous breaking of time-translation symmetry and the emergence of self-sustained oscillations in quantum driven-dissipative systems is the hallmark of continuous time crystals (CTCs). An outstanding challenge is to achieve precise, coherent control of such dynamical phases, whose complex nonlinear dynamics makes them inherently difficult to manipulate. Here, we experimentally demonstrate coherent control of a solid-state CTC realized in a non-resonantly excited spin-polarized exciton-polariton condensate in a Ga(Al)As microcavity. We exploit two complementary control channels: an additional weak control laser and the optomechanical interaction with confined GHz phonons. By tuning the control laser energy and power, we stabilize distinct dynamical regimes, including frequency pushing, injection locking with continuous tuning of the limit-cycle frequency, and full suppression of the autonomous dynamics via phase locking. Under appropriate detuning conditions, the control excitation generates coherent mechanical self-oscillation through the polariton--phonon deformation-potential interaction, evidenced by spectral sidebands. The resulting dynamical back-action provides a phonon-mediated locking channel that fixes the frequency of the CTC. Together, the two channels dramatically enhance the temporal coherence of the GHz limit-cycle dynamics, as established through linewidth narrowing and time-resolved first-order correlation measurements $\gone(\tau)$. Our work establishes a way to harness the unique aspects of CTCs for practical applications in the GHz-range.
		}
	\end{abstract}
	
	\maketitle
	
	%%---------------------------------------------------------------------------%%
	\textit{Introduction.---}Driven-dissipative quantum systems can spontaneously develop self-sustained coherent oscillations (limit cycles) through the interplay of gain, dissipation, and nonlinearity, forming a non-equilibrium state of matter known as a continuous time crystal (CTC), in which time-translation symmetry is broken by the dynamics itself rather than by external periodic modulation~\cite{Wilczek2012,Iemini2018,Zhu2019}. This distinguishes CTCs from discrete time crystals, which arise in periodically driven many-body systems and exhibit subharmonic responses~\cite{Else2016,Khemani2016}. CTCs have been realized in a variety of platforms, including atom-cavity systems~\cite{Kongkhambut2022}, superfluid $^3$He~\cite{Autti2021}, Rydberg gases~\cite{Wu2024}, semiconductor systems~\cite{Greilich2024}, and exciton-polaritons~\cite{Rayanov2015,Nalitov2018, Orfanakis2021}. Recently, we demonstrated a solid-state CTC based on the spinor degree of freedom of an exciton-polariton condensate~\cite{Carraro2024}, where the pseudospin undergoes a self-sustained Larmor-like precession stabilized by the back-action of self-induced GHz phonons.
	
	A central open question is whether the self-sustained, CTC dynamics of driven-dissipative systems can be coherently controlled without losing the underlying self-organized limit-cycle (LC) behavior. Previous approaches for external control of CTCs have been restricted to periodic driving, revealing rich nonlinear phenomena such as synchronization and chaos~\cite{Greilich2025}, and frequency control~\cite{Makinen2025}. Related experiments on polariton condensates have demonstrated enhanced coherence of the oscillation under driven conditions~\cite{Gnusov2023,delValle2023,Yulin2023,Barrat2024,Barrat2024b,Gnusov2025}. However, periodic driving explicitly breaks time-translation symmetry and transforms the dynamics into a forced response, rather than stabilizing the autonomous CTC phase. 
	
	In this Letter, we demonstrate that the temporal dynamics of a CTC can be controlled by a stimulus incommensurate with its intrinsic dynamics. In our case, this is achieved using a weak, energy-tunable continuous-wave control laser~\cite{Gavrilov2016, Gavrilov2018, Leblanc2020, Smirnov2026,Novokreschenov2026,Smirnov2026}, which enables coherent control of a polariton CTC while preserving its self-sustained behavior. The control laser acts directly on the condensate, allowing tuning and stabilization of the CTC oscillation frequency. Under appropriate detuning conditions, it also induces coherent mechanical self-oscillation, whose dynamical back-action provides an additional locking channel. Together, these mechanisms enable precise control of the CTC dynamics and strongly enhance its temporal coherence.

	%%---------------------------------------------------------------------------%%
	\begin{figure*}[ht!]
		\centering
		\includegraphics[width=0.7\textwidth]{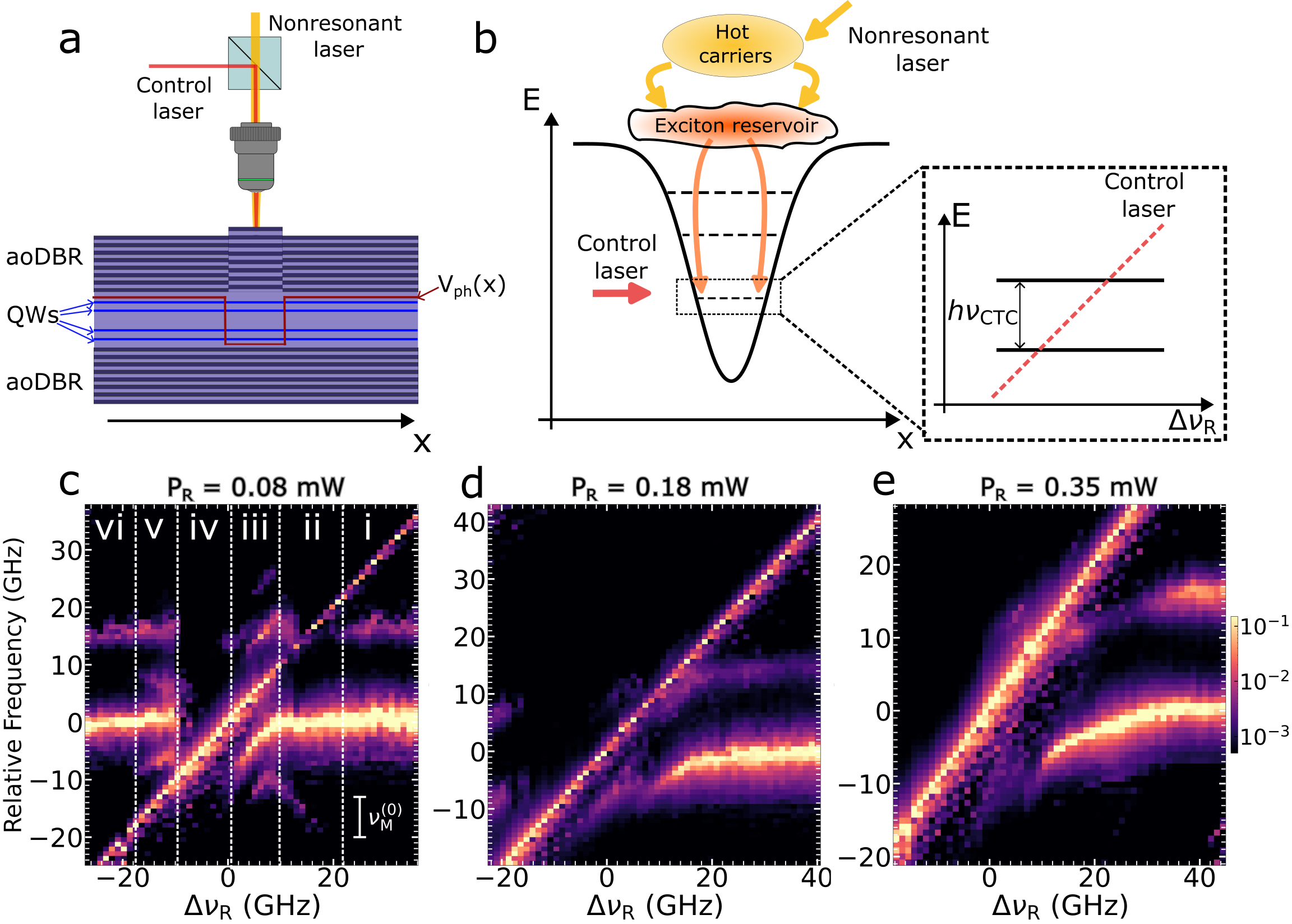}
		\caption{%
			\textbf{Experimental platform and CTC control scheme.}
			(a) A non-resonant and a control cw laser are combined and focused onto a Ga(Al)As planar microcavity with acousto-optic distributed Bragg reflectors (aoDBRs) embedding GaAs quantum wells (QWs). Lateral micropatterning of the cavity spacer defines a photonic and phononic trapping potential $V_\mathrm{ph}(x)$.
			(b) Energy diagram: hot carriers injected by the non-resonant laser form an exciton reservoir that relaxes into the trap, feeding the ground-state polariton condensate in the CTC phase. The control laser (red arrow) is tuned near the ground state. Inset: the CTC doublet (splitting $\hbar\Omctc$) and the control laser energies (dashed red) as a function of detuning $\Delta\nu_\mathrm{R}$.
			(c--e) Emission spectra (logarithmic color scale) as a function of control-laser detuning $\Delta \nu_\mathrm{R}$ from the lower-energy CTC line, for increasing control-laser powers $P_\mathrm{R} = 0.08$, $0.18$, and $0.35$~mW, at fixed non-resonant power $P_\mathrm{NR} = 58$~mW. At the lowest power (c), six dynamical regimes (i--vi) are identified by dashed vertical lines and discussed in the text. Optically locked regimes broaden with increasing $P_\mathrm{R}$.
		}
		\label{Fig1}
	\end{figure*}
	\textit{Platform and time crystal.---}The polariton based CTC-platform exploits the ``double magic coincidence'' of (Al,Ga)As microcavities~\cite{Trigo2002,Fainstein2013, Rozas2014}: a planar distributed acousto-optic Bragg reflector (aoDBR) structure designed for near-infrared photons simultaneously confines GHz longitudinal acoustic phonons, with both fields strongly coupled via the deformation potential interaction mediated by excitons in the embedded GaAs quantum wells (QWs)~\cite{Sesin2023}. The microcavity contains six 15-nm-thick GaAs QWs and has a Rabi splitting of $\sim$6~meV and a quality factor $Q \approx 5000$. Lateral polariton traps are defined by the etch-and-overgrowth method~\cite{Winkler2015,Kuznetsov2018}: a shallow mesa patterned in the cavity spacer (away from the QWs depth) provides a confining potential $V_\mathrm{ph}(x)$ for both photons and phonons simultaneously. Experiments were performed on nominally square traps of $4\times4\,\mu$m$^2$ and $3\times3\,\mu$m$^2$ lateral size, at photon-exciton detunings of approximately $-10$~meV and $-5$~meV, respectively. The confined longitudinal (transverse) acoustic resonances appear at $\nu^{(0)}_\mathrm{M}\sim$7~(5)~GHz and $\nu^{(1)}_\mathrm{M}\sim$21~(14)~GHz~\cite{Kuznetsov2023}. Further sample and fabrication details are given in the Supplemental Material~\cite{SM}.
	
	Measurements were performed at 10~K and 75~K. A continuous-wave (cw) pump Ti:sapphire laser at $\sim$760~nm provides non-resonant excitation of polaritons, focused to a $\sim$5~$\mu$m diameter Gaussian spot on the trap [Fig.~\ref{Fig1}(a)]. A second, much weaker energy-tunable cw control laser (typically $\sim$10$^{-3}$ of the non-resonant pump intensity) is introduced collinearly with the pump and tuned close in energy to the trapped polariton modes, particularly addressing the ground state from which most of the emission originates. Both lasers are linearly polarized. This configuration is illustrated in Fig.~\ref{Fig1}(b). The photoluminescence (PL) is collected and analyzed using either a grating spectrometer (resolution $\sim$0.1~meV) or a voltage-tunable Fabry-P\'{e}rot etalon (resolution $\sim$0.3~GHz) for high-resolution measurements~\cite{Kuznetsov2023,SM}. First-order coherence functions $|\gone(\tau)|$ are obtained using a modified Michelson interferometer with a motorized delay line~\cite{Carraro2024,SM}.
	
	Under non-resonant excitation alone, above a condensation threshold $P_\mathrm{th}$ the system undergoes Bose-Einstein-like condensation of polaritons~\cite{Kasprzak2006,Balili2007}. Upon further increasing the pump power, a spin-bifurcation transition drives the condensate into a spin-polarized state [Fig.~\ref{Fig1}(b)]~\cite{Ohadi2015}. Then, at even higher pump powers, the pseudospin spontaneously enters a self-sustained Larmor-like precession at frequency $\nu_\mathrm{CTC}$, realizing a CTC~\cite{Carraro2024}.
	
	%%---------------------------------------------------------------------------%%
	\textit{Injection locking and phonon-mediated coherent control of the time crystal.---}
	The CTC dynamics is reflected in the emission spectrum, which displays a characteristic doublet associated with the pseudospin precession, with splitting $h\nu_\mathrm{CTC}$. The control laser is tuned with an energy detuning $h\Delta\nu_\mathrm{R} = h(\nu_\mathrm{R} - \nu_0)$ defined with respect to the lower-energy line of this doublet ($h \nu_{0}$). The relative position of the laser with respect to the doublet is illustrated in the inset of Fig.~\ref{Fig1}(b).
	
	Figs.~\ref{Fig1}(c--e) show emission spectra measured as a function of $\DvR$ for a $4\times4\,\mu$m$^2$ trap, in the CTC regime. Each panel corresponds to increasing control laser power $P_\mathrm{R}$ from (c) to (e). In the absence of the control excitation, the emission displays a doublet with splitting $\nu_\mathrm{CTC} \approx 14$~GHz, close to twice the fundamental confined phonon frequency, $\nu_\mathrm{CTC}\sim2\nu^{(0)}_\mathrm{M}$. At the lowest power [panel~(c)], six distinct dynamical regimes are identified.
	
	Starting from positive detuning, in regime~(i) the control laser is far blue-detuned from the CTC doublet and remains essentially uncoupled, leaving the CTC emission unperturbed. As the detuning decreases (ii), the upper CTC line locks to the laser frequency, evidencing optical injection locking~\cite{Chestnov2019, Navarro2022, Arumugam2026}, while the splitting and thus the CTC frequency $\nu_\mathrm{CTC}$ is continuously tuned by $\Delta\nu_\mathrm{R}$. A low-energy sideband mirroring the laser appears due to a parametric process~\cite{Ferrier2010, Zambon2020}. When the laser lies between the two CTC lines (iii) and the detuning approaches the confined phonon frequency $\nu^{(0)}_\mathrm{M} \approx 7$~GHz, both lines follow the laser while maintaining a splitting close to $\nu^{(0)}_\mathrm{M}$, and additional equidistant sidebands emerge. This marks the onset of \textit{optomechanical locking}, where the interaction between the control laser and the CTC lines resonantly drives the confined acoustic mode, whose dynamical back-action on the polariton pseudospin locks the CTC splitting to the mechanical frequency. As the laser approaches the lower CTC line (iv), the doublet collapses into a single narrow line locked to the laser frequency, indicating suppression of the self-sustained limit-cycle and transition to a phase-locked state~\cite{Alonso2024, Beltramo2026}. On the red-detuned side (v), additional sidebands appear due to nonlinear mixing between the CTC and the control excitation~\cite{Chestnov2019}. Finally, for larger detuning (vi), the laser decouples and the CTC doublet recovers its unperturbed frequency and splitting, mirroring regime~(i). The small residual redshift may originate from sample temperature drift during the experiment.
	
	Increasing $P_\mathrm{R}$ from 0.08 mW to 0.35 mW [panels~(d) and~(e)] progressively broadens the optical locking regimes~\cite{Chestnov2019}. For the highest power, regimes~(ii) and~(iv) dominate the scan, with injection locking and CTC collapse persisting over a large fraction of the accessible detuning range. Prior to locking, as the laser approaches each CTC line, they are repelled from the control laser. At a critical detuning, this behavior gives way abruptly to full suppression of the CTC dynamics, evidenced by phase locking of both pseudospin components to the control laser.
	%%---------------------------------------------------------------------------%%
	
\begin{figure}[ht!]
		\centering
		\includegraphics[width=1.0\columnwidth]{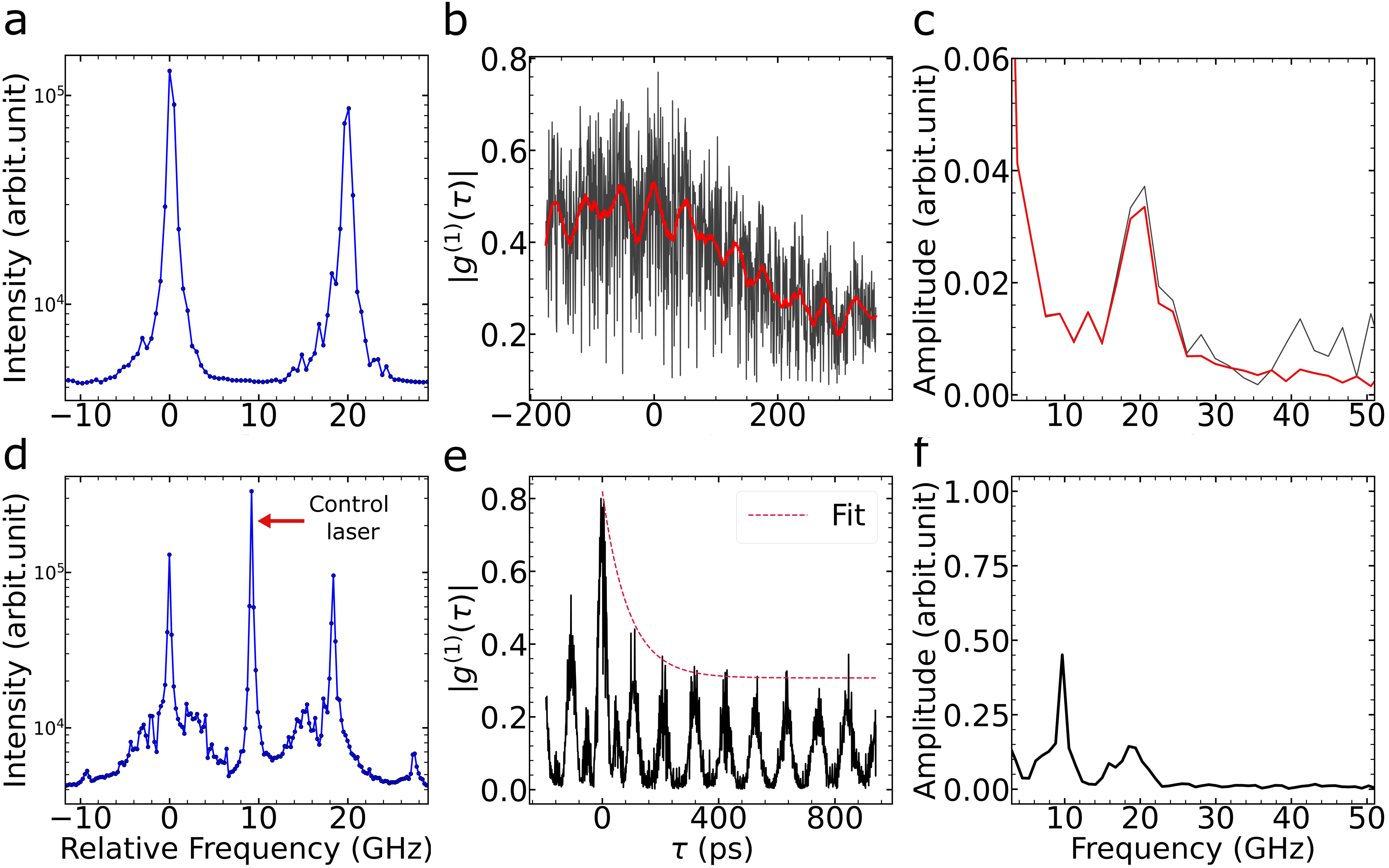}
		\caption{%
			\textbf{Experimental temporal coherence of the CTC with and without the control laser.}
			(a--c) Control laser off.
			(a)~Emission spectrum showing the CTC doublet with splitting $\nu_\mathrm{CTC} \approx 20$~GHz.
			(b)~$|\gone(\tau)|$ showing oscillations at $\nu_\mathrm{CTC}$; red curve: Savitzky--Golay filter.
			(c)~Fourier transform of $|\gone(\tau)|$, normalized to the zero-frequency component, revealing a broad peak at $\nu_\mathrm{CTC}$ with modest amplitude.
			(d--f) Control laser on; arrow in (d) indicates the laser frequency.
			(d)~Emission spectrum displaying a five-line comb with uniform $\sim$10~GHz spacing.
			(e)~$|\gone(\tau)|$ showing persistent oscillations beyond the measurement window; dashed red curve: fit $C + Ae^{-\tau/\tau_0}$ ($C = 0.31$, $\tau_0 = 90$~ps).
			(f)~Fourier transform of $|\gone(\tau)|$, normalized to the zero-frequency component, showing a sharp, resolution-limited peak at $\sim$10~GHz in contrast to the broad low-amplitude peak of panel~(c).
		}
		\label{Fig2}
	\end{figure}

	%%---------------------------------------------------------------------------%%
	\textit{Coherence enhancement.---}Besides enabling coherent control of the CTC dynamics, as demonstrated in Fig.~\ref{Fig1}, the control excitation dramatically enhances the temporal coherence of both the polariton condensate and the CTC. Figures~\ref{Fig2}(a--c) describe the bare CTC in a $3\times3\,\mu$m$^2$ trap without the control laser. The emission spectrum [panel~(a)] displays the characteristic CTC doublet, in this case with a splitting $\nu_\mathrm{CTC} \approx 20$~GHz, locked to the second confined acoustic phonon mode of the cavity, $\nu^{(1)}_\mathrm{M}$. This value differs from the $\sim$14~GHz splitting in Fig.~\ref{Fig1} due to the different trap structure, which modifies the energy splitting of the linearly polarized modes, as well as variations in the excitation conditions affecting the nonlinear interactions~\cite{Ohadi2015}. The first-order correlation function $|\gone(\tau)|$ [panel~(b)] shows oscillations at $\nu_\mathrm{CTC}$, visible after Savitzky--Golay filtering (solid red line). The Fourier transform [panel~(c)] reveals a broad peak (FWHM $\sim$5~GHz, corresponding to a decay time of $\sim$200~ps) at $\nu_\mathrm{CTC} \approx 20$~GHz whose amplitude is modest compared to the zero-frequency component, to which the spectrum is normalized, reflecting the limited coherence of the bare limit-cycle dynamics under these excitation conditions.
	
	We then focus on the representative case in which the control laser lies between the two CTC lines [Figs.~\ref{Fig2}(d--f)], which dramatically transforms the coherence properties. The emission spectrum [panel~(d)] develops a five-line structure: the original doublet at 0 and $\sim$20~GHz, the control laser at $\sim$10~GHz, and additional sidebands at $\sim-10$~GHz and $\sim-30$~GHz, with all lines uniformly spaced by $\sim$10~GHz ($\sim\nu^{(1)}_\mathrm{M}/2$). This regular spacing reflects a subharmonic response of the system, corresponding to an effective halving of the limit-cycle frequency relative to the bare CTC~\cite{Carraro2024, SM}. In addition, a pronounced narrowing of the CTC lines is observed, here limited by the resolution of the Fabry-P\'{e}rot filter. The effect of the control laser on the CTC temporal coherence is striking. The $|\gone(\tau)|$ [panel~(e)] displays two distinct components: a fast-decaying contribution at short delays and a long-lived coherent component that persists without measurable decay beyond the experimental window of $\sim$900~ps. The fast component is associated with the intrinsic polariton dynamics and corresponds to the broad background peaks around $\sim$0 and $\sim$20~GHz in the spectrum of Fig.~\ref{Fig2}(d). In contrast, the long-lived component exhibits oscillations at $\sim$10 GHz (period $\sim$100 ps), reflecting a modified LC dynamics under a control excitation. This two-component structure is directly resolved in the Fourier spectrum [panel~(f)]: a broad peak at $\sim$20~GHz reflects the rapidly decaying contribution, while a sharp, resolution-limited peak at $\sim$10~GHz encodes the long-lived coherent oscillation stabilized by the control excitation. Notably, the 10~GHz peak reaches a $\sim$30-fold amplitude enhancement over the bare CTC Fourier amplitude of $\sim$0.035 [panel~(c)], indicating that the control laser not only extends the temporal coherence but also strongly amplifies the coherent oscillation of the CTC. 
	%This behavior is not restricted to this operating point: coherence enhancement is a general consequence of the resonant excitation across different locking regimes. In particular, the spectral linewidth narrowing of the CTC doublet lines as the laser approaches regime~(iii) of Fig.~\ref{Fig1} is analyzed in the Supplemental Material~\cite{SM}, confirming that the behavior observed in Fig.~\ref{Fig2} is representative of the broader phenomenology.
	
	%%---------------------------------------------------------------------------%%
	\textit{Theoretical model.---}The experiments are modeled by a driven-dissipative spinor Gross--Pitaevskii equation~\cite{Carraro2024,Ohadi2015, Wouters2007} extended to include the coherent control. The polariton spinor components $\psip$, $\psim$ (circular polarization basis) are coupled to spin-resolved exciton reservoirs $\tilde{n}_\sigma$ and to a confined phonon mode of displacement $x$, described by the coupled equations,
	\begin{align}
		i\hbar\frac{d\tilde{\psi}_\sigma}{dt} &= \left(\varepsilon_\sigma + U_0|\tilde{\psi}_\sigma|^2 + U_0^R\tilde{n}_\sigma\right)\tilde{\psi}_\sigma- (J + i\hbar\gamma_d)\tilde{\psi}_{\bar{\sigma}}\nonumber\\
		&\quad + \frac{i\hbar\gamma}{2}(\tilde{n}_\sigma - 1)\tilde{\psi}_\sigma + \tilde{F}_\sigma e^{-i2\pi\nu_\mathrm{R} t} \nonumber\\
		&\quad + \hbar \tilde{\xi}_\sigma(t) \sqrt{\frac{\gamma}{2} (\tilde{n}_\sigma +1)},\label{eq:GP}\\
		\frac{d\tilde{n}_\sigma}{dt} &= \gamma_R\left[p_\sigma - \left(1 + |\tilde{\psi}_\sigma|^2\right)\tilde{n}_\sigma\right],\label{eq:res}\\
		\frac{d^2{x}}{dt^2} &= -\Gamma\frac{dx}{dt} - \omega^2_\mathrm{M} x - 4 \omega^2_\mathrm{M}\lambda_1\,\mathrm{Re}\!\left(\tilde{\psi}_+^*\tilde{\psi}_-\right)
		\nonumber\\
		&\quad - 2 \omega^2_\mathrm{M} \lambda_2 (|\tilde{\psi}_+|^2 + |\tilde{\psi}_-|^2).
		\label{eq:ph}
	\end{align}
	The first term on the RHS of Eq.~\ref{eq:GP} contains the bare polariton energy $\varepsilon_\sigma$, the same-spin polariton--polariton interaction $U_0$, and the polariton--reservoir interaction $U_0^R$. The complex coupling $-(J + i\hbar\gamma_d)\tilde{\psi}_{\bar{\sigma}}$ mixes the two spin components: the real (complex) part $J$ ($\hbar\gamma_d$) arising from the energy (decay rate) difference of the two linearly polarized trap modes~\cite{Carraro2024,Ohadi2015}. The first term in the second line describes stimulated gain from the reservoir and bare polariton decay. The coherent excitation $\tilde{F}_\sigma e^{-i2\pi\nu_\mathrm{R} t}$ acts on both spin components equally ($\tilde{F}_+ = \tilde{F}_- = \tilde{F}$), reflecting the linear polarization of the control laser~\cite{Chestnov2019}. The last term introduces a stochastic noise $\tilde{\xi}_\sigma(t)$, accounting for polariton field fluctuations within a standard Langevin approach to driven-dissipative condensates~\cite{Carusotto2013,Deng2010,Fazio2025}. Equation~(\ref{eq:res}) governs the reservoir dynamics driven by external pump $p_\sigma$ (renormalized with respect to the condensation threshold power) and depleted by decay ($\gamma_\mathrm{R}$) and by stimulated scattering into the condensate. Equation~(\ref{eq:ph}) is a damped harmonic oscillator for the confined phonon: the two driving terms couple to the polariton spinor through the deformation potential interaction, with $\lambda_1$ entering via the Josephson channel $J = J_0 + \lambda_1\omega_\mathrm{M} x(t)$ and $\lambda_2$ via a diagonal energy shift $\varepsilon_\sigma = \varepsilon_{\sigma,0} + \lambda_2\omega_\mathrm{M} x(t)$, where $\omega_\mathrm{M} = 2\pi\nu_\mathrm{M}$~\cite{Reynoso2022,Ramos2024,Carraro2024}. The full parameter set, rescaling, and numerical details are given in the Supplemental Material~\cite{SM}.
	
	Figure~\ref{Fig3}(a) shows the calculated polariton (bottom) and phonon (top) power spectra as a function of $\DR$ (in units of $\Omph$). The phonon colormap brightens when the mechanical mode develops macroscopic self-oscillation. Six dynamical regimes are identified, which correlate with the experimental observations of Fig.~\ref{Fig1}(c). In regime~(1), the control laser is tuned between the two CTC lines, whose splitting is $\sim2\Omph$ (as in the experiment of Fig.~\ref{Fig1}), but with unequal detuning from each, so the interaction remains off-resonant. The resulting dynamics is anharmonic, as seen in the Poincar\'{e} sphere of regime~(1) [Fig.~\ref{Fig3}(b)], where the LC orbit is distorted compared to the undriven CTC (leftmost sphere). In (2) the laser is tuned such that its detuning from each CTC line is $\sim\nu_M$. The phonon spectrum brightens, signaling mechanical self-oscillation \cite{Chafatinos2020}. The dynamical back-action locks the CTC splitting and sidebands to $\Omph$, analogously to regime~(iii) of the experimental Fig.~\ref{Fig1}(c). Because this halves the splitting, leading to period doubling relative to the original limit cycle, the Poincar\'{e} sphere of regime~(2) displays an additional loop compared to the unperturbed trajectory.
	
	Regimes~(3) and~(5) are governed by the nonlinear interaction between the CTC and the control excitation: extra spectral lines appear near the lower CTC line and shift continuously with $\DR$~\cite{Chestnov2019}, mirroring regime~(v) of Fig.~\ref{Fig1}(c). In regime~(4), the laser locks to the lower-energy CTC line, continuously pulling its frequency while leaving the LC topology largely intact, as reflected in the Poincar\'{e} sphere of regime~(4) which retains an orbit close to the undriven case, but with a modified controlled Larmor frequency. This corresponds to the injection-locking behavior of regime~(ii) in Fig.~\ref{Fig1}(c) and behaviors observed in Fig.\ref{Fig1}~(d,e).
	
	Figures~\ref{Fig3}(a,b) were obtained without inclusion of the noise term in Eq.~\ref{eq:GP}. The addition of fluctuations in Figs.~\ref{Fig3}(c,d) reveals the coexistence of broad and narrow spectral features with markedly different coherence times. Figure~\ref{Fig3}(c) compares the emission spectra with (blue) and without (red) the control laser, for parameters corresponding to regime~(2) in panel~(a), where phonon self-oscillation occurs. In the absence of the control laser, the CTC doublet (splitted by $2\Omph$) appears as two broadened peaks. When the control laser is tuned between them, these broad features persist but are dominated by much narrower, high-intensity peaks and additional sidebands separated by $\Omph$, indicating the emergence of a highly coherent component. This behavior is reflected in the first-order correlation function $|\gone(\tau)|$ shown in Fig.~\ref{Fig3}(d): without the control laser, damped oscillations at $2\Omph$ dominate, whereas with the control laser a fast-decaying background (at $2\Omph$) coexists with long-lived oscillations at $\Omph$. This two-component dynamics, combining short- and long-coherence contributions, is very similar to the coherence enhancement observed experimentally in Fig.~\ref{Fig2}. Overall, the theoretical model captures all key experimental observations, including the emergence of distinct dynamical regimes and the phonon-mediated locking under the tunable laser. The excellent qualitative agreement with Figs.~\ref{Fig1} and~\ref{Fig2} confirms that the coupled polariton--phonon dynamics and coherent excitation provide a consistent description of the observed CTC behavior.
	
	%%% FIGURES %%%
	\begin{figure}[ht!]
		\centering
		\includegraphics[width=1.0\columnwidth]{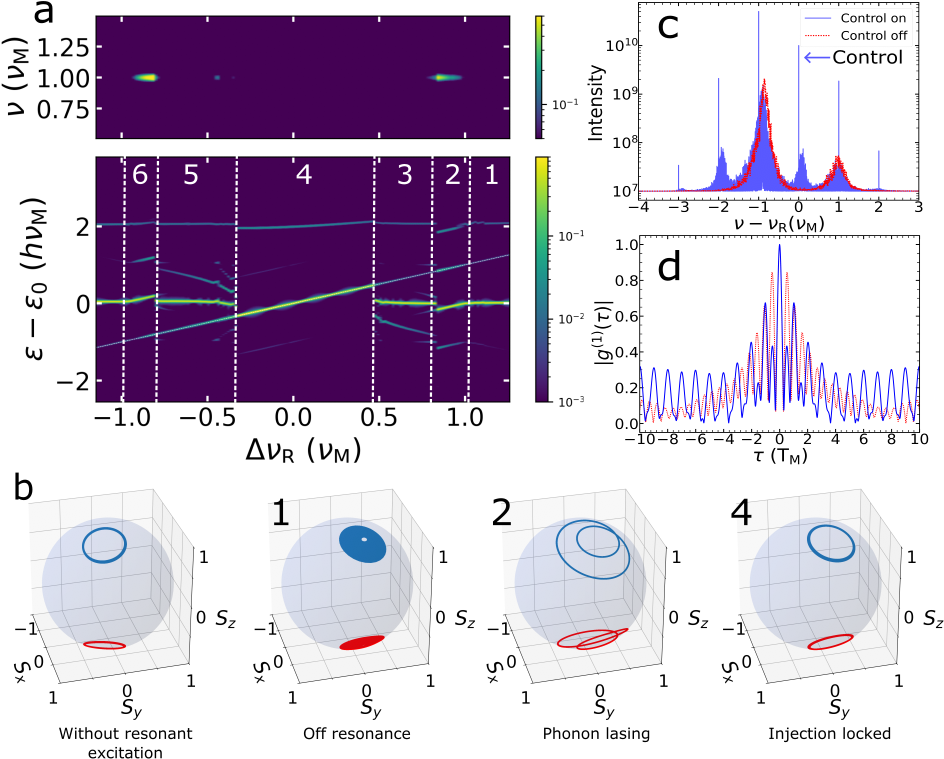}
		\caption{%
			\textbf{Theoretical description of the resonantly driven polariton CTC.}
			(a) Calculated phonon (top) and polariton (bottom) power spectra versus control laser detuning $\DR$ (in units of $\Omph$); enhanced phonon intensity marks the onset of coherent mechanical self-oscillation. Six regimes (1--6), separated by dashed lines, reproduce the experimental phase diagram of Fig.~\ref{Fig1}(c). (b) Pseudospin trajectories on the Poincar\'{e} sphere: undriven CTC (limit cycle), off-resonant (1), phonon-mediated locking (2), and injection locking of the lower-energy line (4). (c) Emission spectra including fluctuations, with (blue) and without (red) the control laser, for parameters in regime (2); narrow, high-intensity peaks and sidebands at $\Omph$ emerge on top of the broadened CTC doublet. (d) Corresponding $|\gone(\tau)|$: without the control (red), damped oscillations at $\sim2\Omph$; with the control laser (blue), a fast-decaying component coexists with long-lived oscillations at $\Omph$, reflecting phonon-mediated locking.
		}
		\label{Fig3}
	\end{figure}
	
	%%---------------------------------------------------------------------------%%
	\textit{Conclusions.---}We have demonstrated coherent control of a solid-state CTC through two complementary channels: direct injection locking of the CTC pseudospin doublet by a control laser, and a phonon-mediated locking channel arising from optomechanical back-action of self-induced coherent mechanical oscillations. By varying the control laser frequency and power, the system can be steered across distinct dynamical regimes, ranging from frequency pulling and partial locking to full spectral collapse, while dramatically enhancing the temporal coherence of the GHz-frequency LC dynamics. A driven-dissipative spinor Gross--Pitaevskii model extended with a coherent excitation term qualitatively reproduces all observations.
	
	The demonstrated control over frequency, coherence, and spectral structure of the CTC opens a route toward coherently controlled polariton oscillators with laser-like coherence, with potential relevance to on-chip frequency references~\cite{TimeCrystalClock}. The combination of injection locking and intrinsic optomechanical coupling explored here provides a general strategy for engineering driven-dissipative CTC phases in solid-state polariton lattices.
	
	%%---------------------------------------------------------------------------%%
	
	%%% END FIGURES %%%
	
	%%---------------------------------------------------------------------------%%
	\begin{acknowledgments}
		This research is funded in part by the Gordon and Betty Moore Foundation, Grant GBMF14019. We also acknowledge partial financial support from the ANPCyT-FONCyT (Argentina) under grants PICT 2019-0371 and PICT 2020-3285. A.S.K. and P.V.S. acknowledge funding from the German DFG (grant 359162958).
	\end{acknowledgments}

	%%---------------------------------------------------------------------------%%

\onecolumngrid

\pagebreak

%\clearpage
\setcounter{section}{0}
\setcounter{page}{1}

%\renewcommand{\thesection}{Supplementary Note \arabic{section}}
%\renewcommand{\figurename}{\textbf{Supplementary Figure} \!\!}
%\titleformat{\section}{\normalfont\Large\bfseries}{S\thesection.}{1em}{}

\setcounter{figure}{0}
\renewcommand{\theHfigure}{A\arabic{figure}}

% --- Supplementary figure numbering ---
\renewcommand{\thefigure}{SM\arabic{figure}}
\renewcommand{\figurename}{Figure}
\newcommand{\FigSM}[1]{Fig.~\ref{#1}}
% --- Supplementary equation numbering ---
\renewcommand{\theequation}{SM\arabic{equation}}

\begin{center}
\textbf{\Large \underline{Supplementary Material for:}\\~\\ \textit{\Large Coherent Control of a Polariton Continuous Time Crystal}}\label{SMat}
\end{center}

%%%%%%%%%%%%%%%%%%%%%%%%%%%%%%%%%%%%%%%%%%%%%%%%%%%%%%%%%%%%
\section{Sample design and experimental details}

\subsection{Sample design}

The microcavity platform is based on the (Al,Ga)As material system, which exhibits a near coincidence between optical and acoustic properties. In particular, the refractive index contrast between the constituent layers closely matches the contrast in acoustic impedance, and the ratios of optical phase velocities closely track those of the acoustic sound velocities across the layers. As a result, the same multilayer structure can simultaneously act as a Bragg reflector for near-infrared photons and GHz longitudinal acoustic phonons. This so-called ``double magic coincidence''~\cite{Trigo2002,Fainstein2013} enables the optimal simultaneous confinement of light and sound within the same geometry.

The hybrid photon--phonon microcavity studied here, including its optical and acoustic properties, has been extensively characterized in previous works~\cite{Kuznetsov2023,Kuznetsov2026}. The planar microcavity (MC) was grown by molecular beam epitaxy on a (001)-oriented GaAs substrate. It consists of an optical cavity spacer with thickness $3\lambda/2$, embedding six 15-nm-thick GaAs quantum wells (QWs), separated by 7.5-nm-thick Al$_{0.1}$Ga$_{0.9}$As barriers. The QWs are positioned close to the antinodes of both the optical electric field and the acoustic displacement field in order to maximize the coupling to confined photonic and phononic modes. The resulting structure exhibits a Rabi splitting of approximately $6$~meV and a cavity quality factor $Q \approx 5000$.

The spacer is sandwiched between upper and lower distributed Bragg reflectors (DBRs). Each DBR period consists of a stack of three pairs of Al$_{x_1}$Ga$_{1-x_1}$As / Al$_{x_2}$Ga$_{1-x_2}$As layers with quarter-wavelength optical thickness, engineered with different aluminum concentrations to optimize simultaneous optical and acoustic reflectivity. This multilayer design produces stop bands for photons with wavelength $\lambda \approx 810$~nm and, simultaneously, for longitudinal acoustic (LA) phonons with wavelengths $\lambda$ and $3\lambda$. For LA phonons propagating along the growth axis, these correspond to frequencies of approximately $21$~GHz and $7$~GHz, respectively, leading to the confinement of discrete acoustic modes within the cavity.

These confined modes exhibit enhanced coupling to the excitonic component of the polaritons via the deformation potential interaction. The operating conditions and trap parameters set the frequency of the pseudospin limit cycle (mode splitting). When this Larmor-like precession frequency approaches either the $\sim 7$ or $\sim 21$~GHz mechanical resonances, the system enters distinct regimes where the limit-cycle dynamics is governed by coupling to different confined phonon modes. This behavior is illustrated in the main text: in Fig.~1(c), the dynamics is dominated by the $\sim 7$~GHz mode, whereas in Fig.~2 the $\sim 21$~GHz mode sets the relevant frequency scale.

\subsection{Lateral confinement: polariton and phonon traps}

Zero-dimensional confinement of both polaritons and phonons is achieved through lateral patterning of the cavity spacer using the etch-and-overgrowth technique~\cite{Winkler2015,Kuznetsov2018}. This approach, which has been employed in previous studies of polariton condensation in the same platform~\cite{Kuznetsov2023,Kuznetsov2026, Kuznetsov2018, Chafatinos2023}, consists of interrupting the molecular beam epitaxy growth after deposition of the spacer region containing the QWs, followed by shallow photolithographic etching of the surface (17~nm). The sample is subsequently reintroduced into the growth chamber for overgrowth of the upper DBR at reduced temperature to preserve the patterned profile.

Due to the conformal nature of the overgrowth, the etched pattern results in local variations of the spacer thickness, forming mesa structures with heights of a few nanometers and lateral dimensions in the micrometer range. These mesas define trapping potentials for both optical and acoustic modes, as the increased spacer thickness in the non-etched regions leads to lower resonance energies compared to the surrounding areas.

In this work, we focus primarily on nominally square traps with lateral size $4 \times 4~\mu\mathrm{m}^2$, operated at a photon--exciton detuning of approximately $\delta_{CX} \approx -10$~meV in the non-etched region. Additional measurements are performed on $3 \times 3~\mu\mathrm{m}^2$ traps with detunings close to $\delta_{CX} \approx -5$ meV, corresponding to excitonic fractions $|X|^2 \approx 0.2$ to $0.05$. The lateral confinement simultaneously quantizes polariton and phonon modes.

The combination of vertical confinement via the DBRs and lateral confinement via the patterned spacer ensures strong spatial overlap between the polariton condensate and the confined acoustic modes~\cite{Chafatinos2023}. Together with the enhanced deformation-potential interaction in the (Al,Ga)As system, this results in efficient coupling between the condensate dynamics and GHz phonons. This hybrid confinement underlies the optomechanical back-action effects and phonon-mediated locking mechanisms discussed in the main text.
%%%%%%%%%%%%%%%%%%%%%%%%%%%%%%%%%%%%%%%%%%%%%%
\begin{figure}[t]
	\centering
	\includegraphics[width=\linewidth]{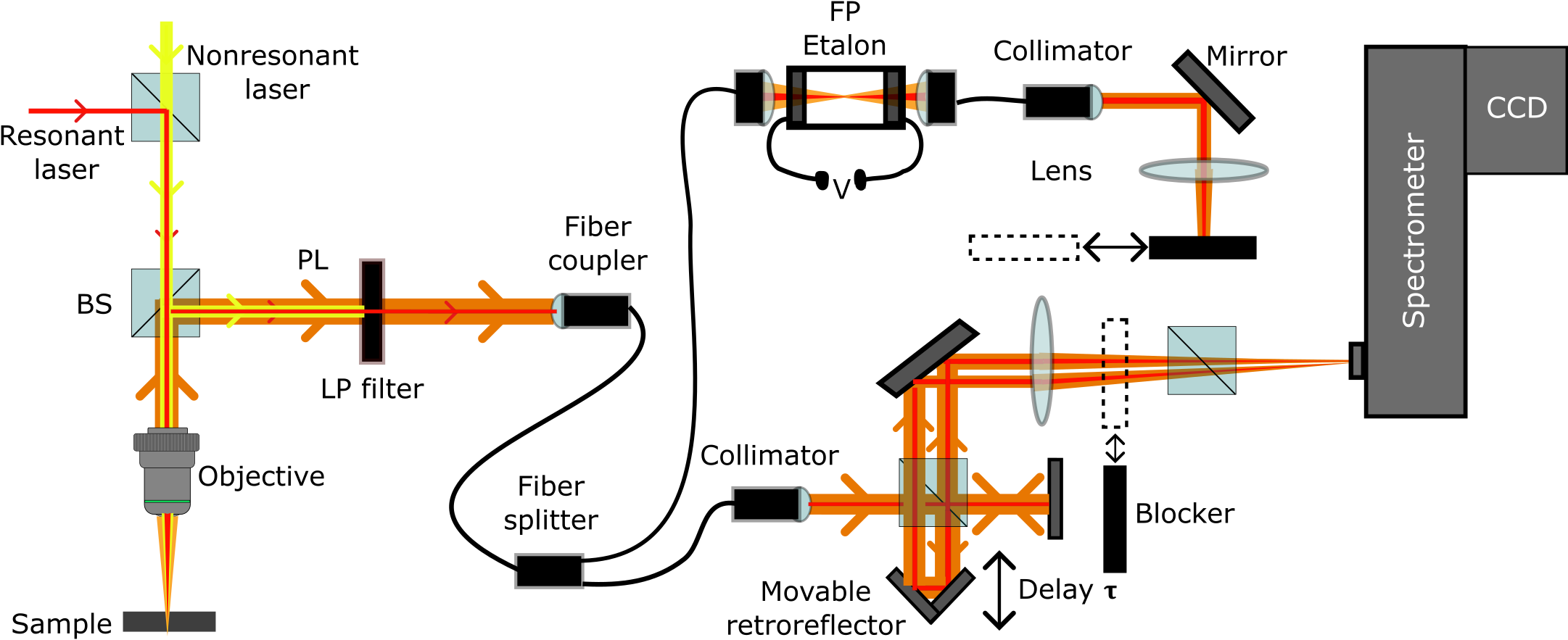}
	\caption{
		\textbf{Experimental setup for excitation, detection, and coherence measurements.}
		A non-resonant continuous-wave Ti:Sapphire laser (yellow, $\sim 760$~nm) and a control energy-tunable diode laser in the near-infrared (red) are combined and focused onto the sample through a microscope objective. The emitted photoluminescence (PL) is collected in reflection geometry, spectrally filtered by a long-pass (LP) filter that suppresses only the non-resonant pump, and coupled into a single-mode fiber. At the fiber output, the signal is split into two branches. One branch is directed to a piezo-tunable Fabry--P\'erot (FP) etalon for high-resolution spectral analysis, followed by a grating spectrometer and CCD detection. The second branch is sent to a Michelson interferometer with a movable retroreflector, introducing a controllable delay $\tau$ between the two arms. The measurement configuration is selected by blocking either branch, as indicated in the figure. This setup enables both high-resolution spectroscopy and measurements of the first-order temporal coherence function $|\gone(\tau)|$ under identical experimental conditions.
	}
	\label{FigS1}
\end{figure}

\section{Experimental setup}

Optical measurements were performed in a liquid-He cryostat at temperatures between 10~K and 75~K. The sample was excited using two continuous-wave lasers in a collinear configuration, as schematically shown in \FigSM{FigS1}.

A strong non-resonant pump laser  (with excitation power tunable in the 30-80 mW range), provided by a continuous-wave Ti:Sapphire source and tuned to $\sim 760$~nm, was used to inject carriers well above the polariton resonance. The beam was focused onto the sample at normal incidence using a microscope objective, resulting in a Gaussian spot with a diameter of $\sim 5~\mu$m positioned on a single trap. The injected carriers relax towards lower energies and populate the confined polariton states, enabling condensation in the ground mode.

In addition, a much weaker control laser, consisting of a tunable diode laser operating in the near-infrared spectral range (typically $\sim 10^{-3}$ of the non-resonant pump power), was introduced collinearly and tuned close in energy to the confined polariton modes. The energy of the laser was precisely tuned using a feedback from a high-resolution wavemeter. This beam selectively addresses the lowest-energy state of the trap, from which most of the emission originates, and provides a controlled coherent excitation on top of the incoherent reservoir created by the non-resonant excitation. The power of the control laser, $P_\mathrm{R}$, is independently controlled.

Both excitation beams are linearly polarized and combined using a beam splitter before being focused onto the sample. The photoluminescence (PL) emitted by the microcavity is collected through the same objective in a reflection geometry. A long-pass (LP) filter is used to suppress the reflected non-resonant pump laser around 760~nm, while transmitting both the control excitation and the cavity emission.

The collected PL is coupled into a single-mode optical fiber. At the fiber output, the signal is directed to different detection paths depending on the measurement configuration. In particular, the emission can be sent either to a high-resolution spectral analysis branch or to an interferometric setup for temporal coherence measurements. The selection between these two configurations is performed by blocking one of the paths, as indicated in \FigSM{FigS1}. This arrangement enables standard spectroscopy, high-resolution spectral measurements, and first-order coherence measurements under otherwise identical experimental conditions.

\subsection{High-resolution spectroscopy with Fabry--P\'erot etalon}

High-resolution spectral measurements are performed using a scanning Fabry--P\'erot (FP) etalon, following the approach developed in Refs.~\cite{Kuznetsov2023, Rozas2014}. This technique enables the resolution of fine spectral features, such as phonon sidebands associated with the coupling between polaritons and confined acoustic modes, as well as signatures of limit-cycle dynamics in the GHz frequency range.

As shown in \FigSM{FigS1}, the collected photoluminescence (PL) is coupled into a single-mode optical fiber and directed to the high-resolution branch of the setup. The signal is first spectrally filtered to suppress residual scattered light from the non-resonant pump and then sent to a piezo-tunable FP etalon. The etalon has a finesse of $\sim 240$ and a free spectral range (FSR) of $68$~GHz, providing sub-GHz spectral resolution.

The transmission of the FP is tuned by applying a voltage to a piezoelectric actuator, allowing controlled scanning of the resonance condition. The transmitted signal is subsequently guided to a grating spectrometer, which separates the different transmission orders of the etalon \cite{Rozas2014}. The resulting spectra are recorded using a nitrogen-cooled CCD camera. A custom acquisition routine is used to control the piezo voltage and reconstruct the high-resolution spectra from the scanned signal.

To ensure stability during the measurements, the FP etalon is actively temperature-stabilized. Because the detection is performed through a single-mode fiber, spatial information is not preserved in this configuration. To avoid collecting emission from multiple traps, all measurements are carried out on isolated traps.

This technique provides direct access to the fine structure of the emission spectrum, enabling the identification of phonon-assisted processes and the determination of the characteristic frequencies of the confined acoustic modes and the condensate self-induced GHz dynamics.

\subsection{First-order temporal coherence measurements}

The first-order temporal coherence function $|\gone(\tau)|$ is measured using a Michelson interferometer, following the approach described in Ref.~\cite{Kuznetsov2026}. This technique provides direct access to the temporal coherence and dynamics of the polariton emission and allows us to characterize coherence properties in both steady-state and dynamically modulated regimes.

The photoluminescence (PL) collected from the sample is coupled into a single-mode optical fiber and directed to the interferometric branch of the setup. At the fiber output, the signal is split into two arms using a 50/50 beam splitter. One arm has a fixed optical path length, while the second arm includes a movable retroreflector mounted on a motorized translation stage, which introduces a controllable time delay $\tau$ between the two paths.

The beams from the two arms are recombined and directed onto the entrance of a spectrometer equipped with a CCD camera, where they interfere. The resulting interferograms are spectrally resolved, allowing us to select the emission from a given polariton mode. For a fixed delay $\tau$, the interference visibility is extracted from the recorded intensity pattern as
\begin{equation}
	V(\tau) = \frac{I_{\max} - I_{\min}}{I_{\max} + I_{\min}},
\end{equation}
which provides a direct measure of the magnitude of the first-order coherence function, $|\gone(\tau)|$.

The temporal coherence is obtained by recording interferograms while scanning the delay $\tau$. This procedure yields the decay and any oscillatory behavior of $|\gone(\tau)|$, which reflects the underlying condensate dynamics. In particular, oscillations in $|\gone(\tau)|$ provide a direct signature of coherent frequency components in the emission, such as those associated with phonon-assisted processes or self-induced limit-cycle dynamics in the GHz range.

All measurements are performed under the same excitation conditions as the spectral measurements, enabling direct correlation between coherence properties and high-resolution spectral features.
%%%%%%%%%%%%%%%%%%%%%%%%%%%%%%%%%%%%%
\section{The model}
\subsection{Renormalization of the polariton equations}

The dynamical equations used in the main text are written in terms of rescaled (dimensionless-density) variables. Here we briefly outline the renormalization procedure starting from the standard generalized Gross--Pitaevskii equations for driven-dissipative polariton condensates, including a stochastic term accounting for fluctuations.

We consider the spinor order parameter $\Psi = (\psi_+,\psi_-)^\mathrm{T}$ and the corresponding exciton reservoir densities $n_\sigma$. Including coherent driving, the equations read
\begin{align}
	i\hbar\frac{d\psi_\sigma}{dt} &= \left(\varepsilon_\sigma + U_1|\psi_\sigma|^2 + U^R n_\sigma \right)\psi_\sigma 
	- (J + i\hbar\gamma_d)\psi_{\bar{\sigma}} \nonumber\\
	&\quad + \frac{i\hbar}{2}(R n_\sigma - \gamma)\psi_\sigma 
	+ F_\sigma e^{-i2\pi\nu_\mathrm{R} t}
	+ \hbar\xi_\sigma(t)\sqrt{\frac{1}{2}\left(Rn_\sigma + \gamma\right)}, \label{eq:GP_noise_dim}\\
	\frac{dn_\sigma}{dt} &= P_\sigma - \left(\gamma_R + R|\psi_\sigma|^2\right)n_\sigma.
\end{align}
Here, $\varepsilon_\sigma$ denotes the bare mode energies, $U_1$ the polariton--polariton interaction strength, and $U^R$ the polariton--reservoir interaction. The term $-(J + i\hbar\gamma_d)$ accounts for coherent (Josephson-like) and dissipative coupling between the two modes. The gain term is governed by the scattering rate $R$, which describes the stimulated transfer of excitons from the reservoir into the condensate, while $\gamma$ and $\gamma_R$ are the decay rates of polaritons and reservoir excitons, respectively. The coherent excitation is described by $F_\sigma e^{-i2\pi\nu_\mathrm{R} t}$.

$\xi_\sigma(t)$ is a complex Gaussian white noise satisfying,
\begin{equation}
	\langle \xi_\sigma(t)\xi_{\sigma'}(t') \rangle = 0,
	\qquad
	\langle \xi_\sigma^*(t)\xi_{\sigma'}(t') \rangle = \,\delta_{\sigma\sigma'}\,\delta(t-t').
\end{equation}
The multiplicative factor reflects fluctuations associated with gain and loss processes, namely polariton decay and scattering from the reservoir. These fluctuations are incorporated and numerically solved using the truncated Wigner approach~\cite{Carusotto2013,Deng2010}.

To simplify the equations and make the relevant scales explicit, we introduce the rescaled variables
\begin{equation}
	\tilde{\psi}_\sigma = \frac{\psi_\sigma}{\sqrt{\rho_0}}, 
	\qquad
	\tilde{n}_\sigma = \frac{n_\sigma}{n_0},
\end{equation}
where the characteristic densities are defined as
\begin{equation}
	n_0 = \frac{\gamma}{R}, 
	\qquad 
	\rho_0 = \frac{\gamma_R}{R}.
\end{equation}
With this choice, $n_0$ corresponds to the reservoir density at which gain compensates polariton losses, while $\rho_0$ sets the natural condensate density scale.

Using these definitions, and introducing the rescaled parameters
\begin{equation}
	U_0 = U_1 \rho_0, 
	\qquad 
	U_0^R = U^R n_0, 
	\qquad
	p_\sigma = \frac{P_\sigma}{P_\mathrm{th}},
\end{equation}
with $P_\mathrm{th} = \gamma \gamma_R / R$ the condensation threshold, the equations become
\begin{align}
	i\hbar\frac{d\tilde{\psi}_\sigma}{dt} &= \left(\varepsilon_\sigma + U_0|\tilde{\psi}_\sigma|^2 + U_0^R\tilde{n}_\sigma\right)\tilde{\psi}_\sigma 
	- (J + i\hbar\gamma_d)\tilde{\psi}_{\bar{\sigma}} \nonumber\\
	&\quad + \frac{i\hbar\gamma}{2}(\tilde{n}_\sigma - 1)\tilde{\psi}_\sigma + \tilde{F}_\sigma e^{-i2\pi\nu_\mathrm{R} t} + \hbar\tilde{\xi}_\sigma(t)\sqrt{\frac{\gamma}{2}\left(\tilde{n}_\sigma + 1\right)}, \label{eq:GP_SM}\\
	\frac{d\tilde{n}_\sigma}{dt} &= \gamma_R\left[p_\sigma - \left(1 + |\tilde{\psi}_\sigma|^2\right)\tilde{n}_\sigma\right].
	\label{eq:res_SM}
\end{align}
Here, $\tilde{F}_\sigma = F_\sigma / \sqrt{\rho_0}$ and $\tilde{\xi}_\sigma = \xi_\sigma / \sqrt{\rho_0}$ are the rescaled coherent control amplitude and Gaussian noise, respectively.

Equations~\eqref{eq:GP_SM} and~\eqref{eq:res_SM} correspond to the form used in the main text [Eqs.~(1)--(2)], and provide a convenient normalization in which densities are expressed relative to the natural gain and loss balance scales of the system.

\subsection{Self-consistent phonon dynamics}

The dynamics of the confined phonon mode is obtained self-consistently from its coupling to the polariton condensate. We consider a single mechanical mode described by the Hamiltonian
\begin{equation}
	H = H_\mathrm{p} + \hbar\omega_\mathrm{M} b^\dagger b + \mathcal{V}(\hat{x}),
\end{equation}
where $H_\mathrm{p}$ describes the polariton degrees of freedom, $b$ ($b^\dagger$) is the phonon annihilation (creation) operator, $\hat{x} = \alpha (b + b^\dagger)$ is the displacement operator, with $\alpha$ a normalization constant, and $\omega_\mathrm{M}$ is the phonon angular frequency, $\omega_\mathrm{M} = 2\pi \nu_\mathrm{M}$. The term $\mathcal{V}(\hat{x})$ describes the optomechanical interaction.

Using the Heisenberg equation of motion and neglecting quantum correlations, the classical displacement $x(t) = \langle \hat{x} \rangle$ satisfies
\begin{equation}
	\frac{d^2 x}{dt^2} + \Gamma \frac{dx}{dt} + \omega_\mathrm{M}^2 x
	= -2 \alpha^2 \frac{\omega_\mathrm{M}}{\hbar} \frac{\partial \mathcal{V}}{\partial x},
	\label{eq:phonon_eom_SM}
\end{equation}
where $\Gamma$ is a phenomenological damping rate.

In our system, the optomechanical interaction arises from the modulation of both the inter-mode coupling and the mode energies by the mechanical displacement. To lowest order, this can be written as
\begin{align}
	\mathcal{V}(\hat{x}) &= \hbar g_1 (b + b^\dagger)\left(a_+^\dagger a_- + a_-^\dagger a_+ \right) \nonumber\\
	&\quad + \hbar g_2 (b + b^\dagger)\left(a_+^\dagger a_+ + a_-^\dagger a_- \right),
\end{align}
where $g_1$ and $g_2$ are coupling constants, and $a_\pm$ denote the polariton mode operators.

Substituting into Eq.~\eqref{eq:phonon_eom_SM}, and using $\hat{x} \propto (b + b^\dagger)$, we obtain
\begin{align}
	\frac{d^2 x}{dt^2} + \Gamma \frac{dx}{dt} + \omega_\mathrm{M}^2 x
	&= -2 \omega_\mathrm{M} g_1 \left(a_+^\dagger a_- + a_-^\dagger a_+ \right) \nonumber\\
	&\quad -2 \omega_\mathrm{M} g_2 \left(a_+^\dagger a_+ + a_-^\dagger a_- \right).
\end{align}

We now return to the classical mean-field approximation by replacing the operators with complex fields, $a_\sigma \rightarrow \psi_\sigma$. Expressing the result in terms of the rescaled fields $\tilde{\psi}_\sigma = \psi_\sigma / \sqrt{\rho_0}$, and absorbing the density scale into the coupling constants, we define the dimensionless parameters
\begin{equation}
	\lambda_i = \frac{g_i \rho_0}{\omega_\mathrm{M}} \quad (i=1,2),
\end{equation}
such that all factors of $\rho_0$ are included in their definition. We then obtain
\begin{align}
	\frac{d^2 x}{dt^2} &= -\Gamma \frac{dx}{dt} - \omega_\mathrm{M}^2 x 
	- 4 \omega_\mathrm{M}^2 \lambda_1 \,\mathrm{Re}\!\left(\tilde{\psi}_+^* \tilde{\psi}_- \right) \nonumber\\
	&\quad - 2 \omega_\mathrm{M}^2 \lambda_2 \left(|\tilde{\psi}_+|^2 + |\tilde{\psi}_-|^2 \right),
	\label{eq:ph_SM}
\end{align}
which corresponds to the phonon equation used in the main text [Eq.~(3)].

Equation~\eqref{eq:ph_SM} describes a damped harmonic oscillator driven self-consistently by the condensate. The two driving terms originate from the deformation potential interaction between polaritons and phonons: the term proportional to $\lambda_1$ arises from the modulation of the inter-mode (Josephson) coupling, $J = J_0 + \lambda_1 \omega_\mathrm{M} x(t)$, while the term proportional to $\lambda_2$ corresponds to a diagonal energy shift of the polariton modes, $\varepsilon_\sigma = \varepsilon_{\sigma,0} + \lambda_2 \omega_\mathrm{M} x(t)$~\cite{Reynoso2022,Ramos2024}. 
%%%%%%%%%%%%%%%%%%%%%%%%%%%%%%%%%%%%%%
\subsection{Parameters}

This subsection summarizes the parameters used in the numerical simulations presented in Fig.~3 of the main text. In the simulations, all dynamical quantities were expressed in units of the confined mechanical mode frequency $\nu_\mathrm{M}$, such that $\nu_\mathrm{M}=1$ in the normalized equations. For the experimental system considered here, this mode corresponds to the confined acoustic resonance near
$\nu_\mathrm{M}=7~\mathrm{GHz}$. Accordingly, the values reported in the last column of Table~\ref{tab:params_SM} are given in units of GHz. The density scales introduced in the renormalization of the polariton equations are
$\rho_0=800$ and $n_0=904$,
which define the characteristic condensate and reservoir populations, respectively. The non-resonant pumping was taken equal for both spin components, $p_+=p_-=1.5$,
corresponding to excitation above the condensation threshold.

\begin{table}[h]
	\centering
	\caption{Parameters used in the simulations of Fig.~3 of the main text. All parameters are expressed in units of the confined phonon frequency $\nu_\mathrm{M}$. The last column gives the corresponding physical values in units of GHz. Energy parameters ($J_0$, $U_0$, $U_0^R$, $\tilde{F}$) are implicitly quoted as energy$/h$.}
	\begin{tabular}{lcc}
		\hline\hline
		Parameter &  $\left[\nu_\mathrm{M}\right]$ & [GHz] \\
		\hline
		$\gamma$        & $0.619$              & $4.33$ \\
		$\gamma_d$      & $-0.143$             & $-1.00$ \\
		$\gamma_R$      & $0.548$              & $3.83$ \\
		$J_0/h$     & $0.405$              & $2.83$ \\
		$U_0/h$     & $0.238$              & $1.67$ \\
		$U_0^R/h$   & $3.333$              & $23.33$ \\
		$\tilde{F}/h$ & $0.0548$           & $0.383$ \\
		$\Gamma$        & $5.95\times10^{-4}$  & $0.00417$ \\
		$\nu_\mathrm{M}$ & $1$              & $7$ \\
		$\lambda_1$     & $1.60\times10^{-4}$  & --- \\
		$\lambda_2$     & $1.60\times10^{-3}$  & --- \\
		$g_1$           & $2.00\times10^{-7}$  & $1.40\times10^{-6}$ \\
		$g_2$           & $2.00\times10^{-6}$  & $1.40\times10^{-5}$ \\
		
		\hline\hline
	\end{tabular}
	\label{tab:params_SM}
\end{table}

These parameters place the system in the regime where the continuous time crystal phase, coherent phonon generation, and injection locking coexist.
%%%%%%%%%%%%%%%%%%%%%%%%%%%%%%%%%%%%%%%%%%%%%%%%%%%%%%%%%%%%
\subsection{Extended theoretical spectra}

\begin{figure}[ht]
	\centering
	\includegraphics[width=0.5\linewidth]{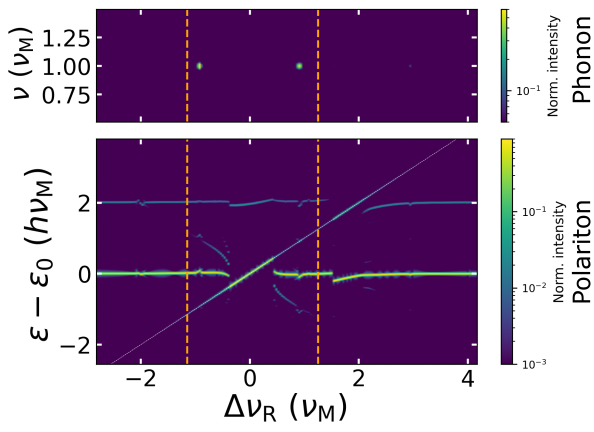}
	\caption{
		\textbf{Extended theoretical spectra.}
		Calculated phonon (top) and polariton (bottom) power spectra as a function of the control laser detuning $\DR$ (in units of $\Omph$), over a wider detuning range. The vertical dashed lines indicate the region shown in Fig.~3(a) of the main text.
	}
	\label{FigS2}
\end{figure}

In order to provide a broader view of the theoretical results discussed in the main text, \FigSM{FigS2} shows the calculated phonon (top) and polariton (bottom) power spectra as a function of the control laser energy detuning $\DR$, over an extended detuning range. This corresponds to the same calculation presented in Fig.~3(a) of the main text. The region displayed in the main text is indicated here by vertical dashed lines, which delimit the range where the most relevant dynamical features occur. This zoomed region was selected to clearly resolve the onset of optomechanical (OM) effects and the different dynamical regimes of the condensate, which would otherwise appear compressed on the full scale.

The extended view highlights the global structure of the spectrum and allows one to track the evolution of the polariton branches over a wider detuning range. In analogy with the experimental observations shown in Fig.~1(d,e) of the main text, similar behaviors are recovered. As the control excitation approaches the upper-energy branch of the time-crystal doublet, this branch is repelled and shifts away from the laser frequency. Upon further tuning, once injection locking is established with the upper branch, the lower-energy branch follows the laser and shifts towards lower energies with decreasing detuning.

These features confirm that the theoretical model captures the essential mechanisms underlying the experimentally observed dynamics, including the interplay between coherent driving, nonlinear interactions, and phonon-mediated processes.

\subsection{Sub-harmonic phonon excitation}

\begin{figure}[ht]
	\centering
	\includegraphics[width=0.7\linewidth]{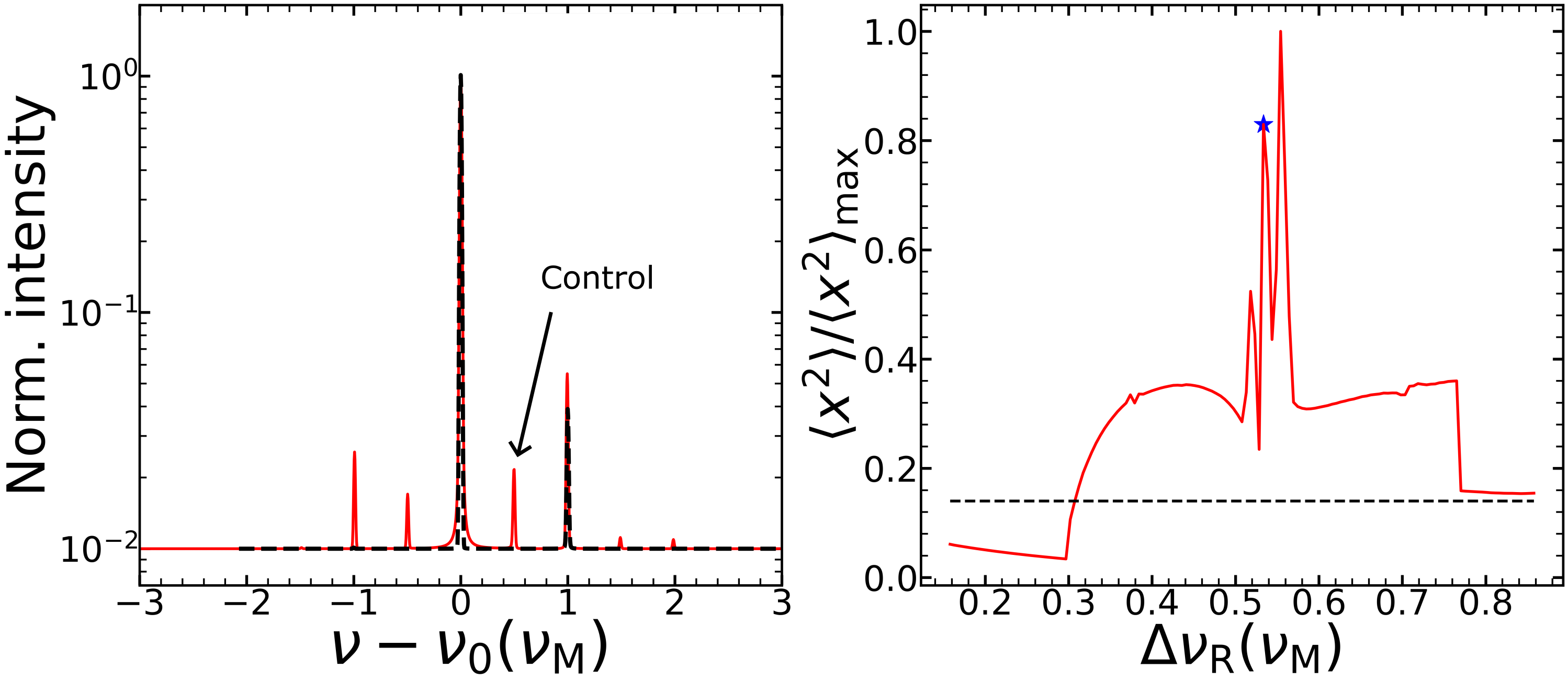}
	\caption{
		\textbf{Sub-harmonic phonon generation.}
		Left: calculated polariton emission spectra for a detuning where the control laser lies between the two time-crystal spectral branches. The dashed black curve corresponds to the undriven case, while the red curve includes the coherent excitation. Additional sidebands emerge with spacing $\nu_\mathrm{M}/2$, indicating frequency locking of the condensate oscillation to half the mechanical frequency. Right: normalized phonon population $\langle x^2\rangle/\langle x^2\rangle_{\max}$ as a function of control laser detuning $\Delta\nu_\mathrm{R}$. The dashed horizontal line marks the undriven reference level. The blue star indicates the detuning corresponding to the spectrum shown in the left panel.
	}
	\label{FigS3}
\end{figure}

In addition to the integer-frequency locking regimes discussed in the main text, the model also supports a subharmonic synchronization regime when the control laser is tuned between the two spectral branches of the continuous time crystal. This situation is analogous to the experimental configuration of Fig.~2 of the main text, where the intrinsic doublet splitting is approximately $20$~GHz and coherent driving near the midpoint leads to linewidth narrowing together with the appearance of sidebands separated by $\sim 10$~GHz.

The theoretical counterpart is shown in \FigSM{FigS3}. In the absence of coherent driving (black dashed curve in the left panel), the emission displays the two natural branches of the self-oscillating condensate. When the control laser is applied at an intermediate frequency (red curve), new spectral sidebands appear with spacing $\nu_\mathrm{M}/2$. This demonstrates that the coherent perturbation can entrain the condensate dynamics so that the oscillation frequency becomes locked to half of the phonon frequency.

The right panel shows the corresponding phonon population, quantified through the normalized $\langle x^2\rangle$, as a function of the laser detuning. Over a finite interval centered between the two branches, the phonon amplitude increases substantially above the undriven baseline (dashed line), revealing enhanced self-consistent mechanical excitation. The blue star marks the operating point used for the spectrum displayed in the left panel.

These results indicate that coherent control excitation can stabilize not only direct synchronization to the mechanical mode, but also fractional locking regimes in which the condensate oscillation reorganizes at a subharmonic of the phonon resonance. This provides a natural explanation for the half-frequency sidebands observed experimentally.

%%%%%%%%%%%%%%%%%%%%%%%%%%%%%%%%%%%%%%%%%%%%%%%%%%%%%%%%%%%%%%%%%%%%%%%%%%%%%%%%%%%%%%%%%%%%%%%%%%%%%%%%%%%%%%%%%%%%%%%%%%%%%%%%%%%%%%%%%%%%%%%%%%%%%%
%%%%%%%%%%%%%%%%%%%%%%%%%%%%%%%%%%%%%%%%%%%%%%%%%%%%%%%%%%%%%%%%%%%%%%%%%%%%%%%%%%%%%%%%%%%%%%%%%%%%%%%%%%%%%%%%%%%%%%%%%%%%%%%%%%%%%%%%%%%%%%%%%%%%%%
\onecolumngrid  % Switch to a single-column layout

\end{document}